# Numerical simulation of flow instability and heat transfer of natural convection in a differentially heated cavity

Hua-Shu Dou[*1], Gang Jiang[2]

[1]Faculty of Mechanical Engineering and Automation,
Zhejiang Sci-Tech University, Hangzhou, China

[2]Huadian Electric Power Research Institute, Hangzhou, China

*Corresponding author:   Email: huashudou@yahoo.com

Abstract: This paper numerically investigates the physical mechanism of flow instability and heat transfer of natural convection in a cavity with thin fin(s). The left and the right walls of the cavity are differentially heated. The cavity is given an initial temperature, and the thin fin(s) is fixed on the hot wall in order to control the heat transfer. The finite volume method and the SIMPLE algorithm are used to simulate the flow. Distributions of the temperature, the pressure, the velocity and the total pressure are obtained. Then, the energy gradient theory is employed to study the physical mechanism of flow instability and the effect of the thin fin(s) on heat transfer. Based on the energy gradient theory, the energy gradient function K represents the characteristic of flow instability. It is observed from the simulation results that the positions where instabilities take place in the temperature contours accord well with those of higher K value, which demonstrates that the energy gradient theory reveals the physical mechanism of flow instability. Furthermore, the effects of the fin length, the fin position, the fin number, and Ra on heat transfer are investigated. It is found that the effect of the fin length on heat transfer is negligible when Ra is relatively high. When there is only one fin, the most efficient heat transfer rate is achieved as the fin is fixed at the middle height of the cavity. The fin blocks heat transfer with a relatively small Ra, but the fin enhances heat transfer with a relatively large Ra. The fin(s) enhances heat transfer gradually with the increase of Ra under the influence of the thin fin(s). Finally, a linear correlation of Kmax with Ra is obtained which reveals the physical mechanism of natural convection from different approaches.

**NOMENCLATURE**

$A$ aspect ratio of the cavity (dimensionless)

$\bar{A}$ amplitude of the disturbance distance     m

$c_p$ specific heat     $m^2\, s^{-2}\, K^{-1}$

$E$ total mechanical energy per unit volumetric fluid     $J\, m^{-3}$

$F$ function (dimensionless)

$g$ gravity acceleration     $m\, s^{-2}$

$H_1$ height of the cavity     m

$H$ loss of total mechanical energy     $J\, m^{-3}$

$k$ thermal conductivity factor     $W\, m^{-1} K^{-1}$

$K$ dimensionless function expresses the ratio of transversal energy gradient and streamwise energy gradient

$l$ length of fin     m

$L$ length of cavity     m

$n$ coordinate in transverse direction     m

Nu Nusselt number along a vertical straight line (dimensionless)

$p$ pressure     $Nm^{-2}$

$p_0$ total pressure     $Nm^{-2}$

Pr Prandtl number (dimensionless)

Ra Rayleigh number (dimensionless)

$s$ coordinate in streamwise direction     m

$t$ time     s

$T$ temperature     K

$T_0$ temperature of the fluid     K

$T_c$ temperature of the cold wall     K

$T_h$ temperature of the hot wall     K

$u, v$ velocity components in x, y directions respectively     $m\, s^{-1}$

$v_m'$     amplitude of the disturbance of velocity in transverse direction     $m\, s^{-1}$

$x, y$ coordinates     m

*Greek symbols*

$\beta$ coefficient of thermal expansion     $(10^{-6})K^{-1}$

$\lambda$ thermal conductivity     $W\, m^{-1}\, K^{-1}$

$\mu$ dynamic viscosity     $Nm^{-2}\, s$

$\nu$ kinematic viscosity     $m^2 s^{-1}$

$\rho$    density of fluid    kg m$^{-3}$

$\omega_d$    frequency of the disturbance    rad s$^{-1}$

$\Delta E$    energy difference along transverse direction    J m$^{-3}$

$\Delta H$    energy difference along streamwise direction    J m$^{-3}$

$\Delta T$    temperature difference    K

$\Delta x$    grid mesh sizes    m

# 1. Introduction

Transient natural convection flows in a cavity are common in industrial applications such as in heat exchangers, solar collectors and nuclear reactors etc, and in our daily life such as in light emitting diode (LED) street lights, computers, mobile phones etc. In the early stages, the steady-state flow has been extensively explored in Refs. [1-3]. Actually, most buoyancy-driven flows in nature and industrial applications are unsteady, and consequently more and more experimental and numerical studies are gradually focusing on unsteady-state flows [4-5]. Patterson and Imberger [6] studied theoretically the transition of unsteady natural convection in a rectangular cavity and found that the whole base flow during the transition includes a vertical boundary layer, a horizontal intrusion and the flow in the core.

The transition of natural convection flows in the cavity differentially heated has been given considerable attention over the last three decades [7-9], and more practical applications based on transient natural convection heat transfer face the problem that it is difficult to enhance or depress the heat transfer rate. It is known that the heat transfer rate will be enhanced when the base flow in the cavity loses its stability [10]. Based on the previous research results, one of the simplest ways to control the heat transfer rate is to fix a solid block on either the hot or the cold side of the differentially heated cavity.

Xu et al. made their efforts to investigate the physical mechanical of natural convection by imposing a thin fin on the hot side of the differentially heated cavity [11-15]. In the following description, we will briefly review the investigation progress and the relevant conclusions. Xu et al. [11] made some experiments to validate the transition of unsteady natural convection using a shadowgraph technique and measured the convection phenomenon using fast-response thermistors. They observed that the transition from initiation by suddenly heating to a quasi-steady state undergoes a number of stages as analyzed by Patterson and Imberger [6]. They also observed separation and oscillations of the thermal flow above the fin and demonstrated that these oscillations trigger instability of the downstream thermal boundary layer flow and enhance the convection. Based on the related experimental results, Xu et al. [12] numerically studied the transition to a periodic flow induced by a thin fin. They find that in the early stage of the flow development following suddenly heating, a lower intrusion front is formed under the fin, and a starting plume arises after the lower intrusion front bypasses the fin. The starting plume induces strong perturbations and even turbulence in the downstream vertical boundary layer. They confirmed that the presence of the thin fin changes the flow regime by triggering intermittent plumes at the leeward side which in turn enhances heat transfer rate. Later, Xu et al. [13] made some further research of transient natural convection flows around a thin fin by both scaling analysis and direct numerical simulations. Both results indicate that the thickness and velocity of the transient natural convection flows around the fin are determined by different dynamic and energy balances which depend on the Rayleigh number, the Prandtl number and the fin length. Furthermore, Xu et al. [14] still performed some experiments to measure the temperature of the thermal flows. They clarify that the oscillatory property of the boundary layer is a key to developing processes by which the boundary layer may be triggered into transition to turbulence, and the consequent enhancement of the total heat transfer. Recently, Xu et al. [15] numerically investigated the relationship between the conductivity of the thin fin and the heat transfer rate of the unsteady natural convection flow adjacent to the finned sidewall of a differentially heated cavity. They observe that the conducting fin improves the transient convection flows in the cavity and enhances heat transfer by up to 52% in comparison with the case without a thin fin.

During the same period, other researchers made lots of contribution in investigation of transient natural convection. Bilgen [16] carried out a numerical study in differentially heated cavities using following parameters: Rayleigh number from $10^4$ to $10^9$, dimensionless thin fin length from 0.1 to 0.9, dimensionless thin fin position from 0 to 0.9, dimensionless conductivity ratio of thin fin from 0 to 60. He finds that Nusselt number is an increasing function of Rayleigh number, and a decreasing function of fin length and relative conductivity ratio. Moreover, the heat transfer may be suppressed by up to 38% by choosing appropriate thermal and geometrical fin parameters. Oztop and Bilgen [17] numerically investigated the heat transfer in a differentially heated, partitioned, square cavity containing heat generating fluid. They clarify that the flow field was modified considerably with partial dividers and heat transfer was generally reduced particularly when the ratio of internal and external Rayleigh number is from 10 to 100. Mezrhab et al. [18] studied the heat transfer in an inclined square cavity, differentially heated by using numerical coupling between the Lattice-Boltzmann equation and finite-difference for the temperature. He finds that when Ra is low ($Ra \leq 10^5$), the average hot wall Nusselt number is higher in inclined cavities than in vertical one; while at large Ra ($Ra = 10^6$), the opposite phenomenon occurs. Varol et al. [19] numerically analyzed the natural convection in solid adiabatic thin fin attached to porous right triangular enclosures. They find that the thin fin can be used as a passive control element for flow field, temperature distribution and heat transfer. Mahmoudi et al. [20] numerically investigated the natural convection cooling of a heat source horizontally attached to the left vertical wall of a cavity filled copper-water. They observe that the increase of Rayleigh numbers strengthens the natural convection flows which leads to the decrease in heat source temperature.

As mentioned above, the researchers have handled so many ways to control the heat transfer mainly by triggering the base flow to lose its stability [6, 11-20]. However, the dynamic mechanisms of flow instability of natural convection around a fin and the relation between instability and heat transfer rate is still not fully understood. These observations motivate the study in this paper.

After almost 20 years of investigation, Dou and co-authors [21-29] proposed an energy gradient theory which describes the rules of fluid material stability from the viewpoint of mechanical energy field. It is claimed that the instability of material system could not be resolved by Newton's three laws, for the reason of a material system moving in some cases is not simply due to the role of forces, but due to the gradient of the total mechanical energy. This approach explains the mechanism of flow instability from physics and derives the criteria of turbulent transition. Accordingly, this method dose not attribute Rayleigh-Benard problem to forces exerted on fluid, but to the gradient of total mechanical energy of fluid. It postulates that when the fluid is placed on a horizontal plate and it is heated from below, the fluid density in the bottom becomes low which leads to mechanical energy gradient $\partial E/\partial y > 0$ along y-coordinate. Only when $\partial E/\partial y$ is larger than a critical value, will the flow become unstable and then fluid cells of vortices will be formed. In addition, the calculating results with the energy gradient theory are in agreement with the experimental data of pipe Poiseuille flow, plane Poiseuille flow, plane Couette flow, Taylor-Couette flow and so on [22-29]. And recently, Dou et al. [30] applied the energy gradient theory into natural convection to reveal the physical mechanism of flow instability in an inclined cavity. The good accordance between the criteria of flow instability based on energy gradient theory and the numerical results indicates that energy gradient theory can reveal the physical mechanism of flow instability in an inclined cavity.

In this study, the energy gradient theory will be used to investigate the unsteady natural convection and to study the physics mechanism of flow instability in a differentially heated cavity. Effect of geometry and flow parameters on heat transfer in natural convection will be discussed. Further, the correlation of the energy gradient function and Rayleigh number in describing natural convection will be analyzed.

# 2 Computational geometry and numerical procedures

## 2.1 Computational geometry

The computational geometry, which is referred to Xu et al. [12], is shown in Fig. 1. The origin of the coordinates is at the lower left corner of the cavity. The length of the cavity is L=1m, and the height of the cavity is $H_1$ =0.24m. Here l expresses the fin length, which is a variable. Furthermore, the thin fin, which is adiabatic, is fixed at the heated side of the cavity, and the fin thickness is considered small or negligible in comparison with the influence of the fin length on heat transfer. $T_0$ is the initial temperature of the unsteady base flow in the cavity, in which water medium is filled. The Prandtl number Pr of water is 6.67 as in Ref. [12]. Initially, the cold wall is cooled to $T_0 - \Delta T$, while the hot side is heated to $T_0 + \Delta T$. The other boundary conditions and initial conditions are described as follows:

X=0: u=v=0, $T_c = T_0 - \Delta T$;

X=L: u=v=0, $T_h = T_0 + \Delta T$;

Y=0: u=v=0, $\partial T / \partial y = 0$;

Y=W: u=v=0, $\partial T / \partial y = 0$.

**2.2 Governing equations**

The development of natural convection in a differentially heated cavity is governed by the following continuity equation, two-dimensional Navier-Stokes and energy equations, and these equations are based on Boussinesq approximation,

$$\frac{\partial u}{\partial x} + \frac{\partial v}{\partial y} = 0 \tag{1}$$

$$\frac{\partial u}{\partial t} + u\frac{\partial u}{\partial x} + v\frac{\partial u}{\partial y} = -\frac{1}{\rho}\frac{\partial p}{\partial x} + \upsilon(\frac{\partial^2 u}{\partial x^2} + \frac{\partial^2 u}{\partial y^2}) \tag{2}$$

$$\frac{\partial v}{\partial t} + u\frac{\partial v}{\partial x} + v\frac{\partial v}{\partial y} = -\frac{1}{\rho}\frac{\partial p}{\partial y} + \upsilon(\frac{\partial^2 v}{\partial x^2} + \frac{\partial^2 v}{\partial y^2}) + g\beta(T - T_0) \tag{3}$$

$$\frac{\partial T}{\partial t} + u\frac{\partial T}{\partial x} + v\frac{\partial T}{\partial y} = k(\frac{\partial^2 T}{\partial x^2} + \frac{\partial^2 T}{\partial y^2}) \tag{4}$$

where $x$ and $y$ are the horizontal and vertical coordinates with origin at the lower left corner of the cavity, $t$ is the time, $T$ is the temperature, $p$ is the pressure, $u$ and $v$ are the velocity components in the $x$ and $y$ directions, $g$ is the acceleration due to gravity, $\beta$ is the coefficient of thermal expansion, $\rho$ is the fluid density, $k$ thermal conductivity factor and $\nu$ is the kinematic viscosity.

## 2.3 Numerical algorithm

The governing equations (1)-(4) are implicitly solved using a finite-volume SIMPLE scheme, with the QUICK scheme approximating the advection term. The diffusion terms are discretized using central differencing with second order accurate. A second order implicit time-marching scheme has also been used for the unsteady term. The discretized equations are iterated with specified under-relaxation factors. In addition, it should be noted that the flow is unsteady and all the contours of variables are time dependent.

In the study, the heat transfer rate can be quantitatively measured by the Nusselt number (Nu) along the cold wall. Here, Nu is a mean value of local Nusselt number along a vertical straight line. The Nusselt number Nu can be expressed as follow:

$$Nu = \frac{1}{\lambda(2\Delta T)/L} \cdot \frac{1}{H} \int_0^H \left| \frac{\rho \cdot C_p}{\lambda} u \cdot (T - T_0) - \frac{\partial T}{\partial x} \right|_x dy \qquad (5)$$

Here, $\lambda$ is the thermal conductivity. $\Delta T$ is the temperature difference, and $c_p$ is the specific heat capacity.

# 3 Validation of the numerical algorithm and grid independence test

### 3.1 Validation of the numerical algorithm

Figure 2 shows the numerical results and experimental results at the same time achieved by Xu et al. in Ref. [12]. Figure 3 shows the present numerical simulation results with the same fluid material, boundary conditions, initial conditions and governing parameters as in Ref. [12]. Comparing the present results with the results in Fig. 2, it is easy to observe that the present results accord well with the results achieved in Ref. [12], and this excellent agreement in turn demonstrates that the numerical algorithm employed in this paper is reliable. Furthermore, in the following study, air will be chosen as the fluid material in the simulation of natural convection. In order to have a quantitatively validation of the computational method, comparison of the positions of the thermal flow plume fronts on time is shown in Table 1. It is found that the simulations agree well with the experiment and previous simulations in reference [12].

**3.2 Grid independence test**

It is known in Ref. [31] that with the flow time starts from 0 second, the subsequent development of natural convection in the cavity is determined by three governing parameters: the Rayleigh numbers Ra, the Prandtl number Pr and the aspect ratio A. And these parameters are defined as follows: $Ra = g\beta\Delta TH^3/\nu k$, $\Pr = \nu/k$, $A = H_1/L$, and the Prandtl number of air is 0.67 in this study. Figure 4 shows the temperature contours at t=60s with four different grid systems, in which grid (A) is meshed with $\Delta x = 0.004$, grid (B) is meshed with $\Delta x = 0.003$, grid (C) is meshed with $\Delta x = 0.002$, and grid (D) is meshed with $\Delta x = 0.0015$, and the Rayleigh number is $3.38\times 10^9$, which is the largest one in this study. Note that all the boundary conditions, the initial conditions, the above governing parameters and time step are all kept the same. It is quantitatively observed from Fig. 4 that all the numerical results are the same. Figure 5 plots the temperature versus time at the same point in four different mesh sizes. And the highest temperature difference of the four different mesh sizes is nearly 0.8 degree, which is negligible in comparison to the temperature difference of the hot and cold wall. And the trends of the temperature variation in the four grid models are almost the same. It should be mentioned that when the mesh is generated, the aspect ratio of the mesh is almost the same for different mesh sizes. Thus, the mesh independence is shown simultaneously both in x and y directions. It is easy to make a conclusion from Figs. 4-5 that the numerical result is independent of the grid size in the studied range.

# 4 Application of energy gradient theory

**4.1 Energy gradient theory**

It is observed in Fig. 4 that the base flow loses its stability by forming a horizontal intrusion and an uprising plume, and it is helpful to enhance heat transfer. However, the physics mechanism of the flow instability is still fully understood. In the following, the energy gradient theory will be applied to explain the physics mechanism of flow instability in natural convection.

From the classical theory of Brownian motion, the fluid particles exchange energy and momentum all the time via collisions. The fluid particle will collide with other particles in transverse directions as it flows along its streamline, and this particle would obtain energy expressed as $\Delta E$ after many cycles. At the same time, the particle would drop energy due to viscosity friction along the streamline. With the same periods, the energy loss expressed as $\Delta H$ would be considerable. Consequently, there exists a critical value of the ratio of $\Delta E$ and $\Delta H$, above which the fluid particle would leave its equilibrium by moving to a new streamline with higher energy or lower energy, and below which the particle would not leave its streamline for its oscillation would be balanced by the viscosity role along the streamline. The criterion of stability can be expressed as follows as in [22-29],

$$F = \frac{\Delta E}{\Delta H} = \left(\frac{\partial E}{\partial n}\frac{2\bar{A}}{\pi}\right) \bigg/ \left(\frac{\partial H_s}{\partial s}\frac{\pi}{\omega_d}u\right) = \frac{2}{\pi^2}K\frac{\bar{A}\omega_d}{u} = \frac{2}{\pi^2}K\frac{v'_m}{u} < Const \qquad (6)$$

where,

$$K = \frac{\partial E/\partial n}{\partial H/\partial s} \qquad (7)$$

Here, u is the streamwise velocity of main flow, $\bar{A}$ is the amplitude of the disturbance distance, $\omega_d$ is the frequency of the disturbance, $v_m' = \bar{A}\omega_d$ is the amplitude of the disturbance of velocity, and $\pi$ is the circumference ratio. Furthermore, F is a function of coordinates which expresses the ratio of the energy gained in a half-period by the particle and the energy loss due to viscosity in the half-period. K is a dimensionless field variable (function) and means the ratio of transversal energy gradient and the rate of the energy loss along the streamline. $H$ is the loss of the total mechanical energy per unit volumetric fluid along the streamline for finite length, which is derived from the Navier-Stokes equations. $E = p + 1/2\rho V^2$ expresses the total mechanical energy per unit volumetric fluid for incompressible flow, s is along the streamwise direction and n is along the transverse direction.

**4.2 Criterion of instability based on energy gradient theory in natural convection**

In the present study, the fluid fills in a differentially heated cavity and the base flow is initially stationary. According to the energy gradient theory [22-29], the stability of a flow depends on the ratio of the gradient of the total mechanical energy in transverse direction and the loss of the total mechanical energy in streamwise direction. For the fluid in a differentially heated cavity, the energy loss is low due to low velocity and its influence is not large. Thus, the nominator in Eq.(7) can represent the feature of extent of flow stability. Based on the particular condition of the base flow, the magnitude of the gradient of the total mechanical energy in natural convection can be written as:

$$K = \sqrt{(\partial E/\partial x)^2 + (\partial E/\partial y)^2} \qquad (8)$$

It should be noted that the influence of gravity is neglected in this study, then we will get $E \sim p_0$, consequently, $K = \sqrt{(\partial p_0/\partial x)^2 + (\partial p_0/\partial y)^2}$, which is shown in Fig. 6. Here, $p_0$ represents the total pressure. From above discussion, the magnitude of the gradient of the total mechanical energy in Eq.(8) can be employed to analyze the instability in natural convection.

# 5 Results and discussions

**5.1 Physical mechanism of flow instability**

Patters and Imberger [6] observed that the flow instability could enhance heat transfer rate of natural convection, however, the physical mechanism of flow instability of natural convection needs to be further clarified. Here, we will explain the physical mechanism of flow instability in natural convection by using the energy gradient theory.

Figure 7(a) shows the contours of the temperature and Figure 7(b) shows the contours of the value of K at 10s with $Ra = 3.38 \times 10^9$. It can be seen in Fig. 7(a) that the base flow downstream of the fin loses its stability by forming small vortices and by moving the fluid in a wave form. Based on the criterion of flow instability of energy gradient theory, there is a critical value of K, above which the flow will lose its stability. It can be seen in Fig. 7(b) that there are some peak areas in the K contour, and flow instability could most likely occur in these peak areas based on the former analysis. In other words, the position with a higher energy gradient function K is easier to lose its stability based on energy gradient theory. By comparing Fig. 7(b) to Fig. 7(a), it is amazedly found that the positions where instabilities take place are in accordance with the area with the higher value of K. The reason why the base flow loses its stability is that the value of the energy gradient function at some positions exceeds its critical value in a flow field. This result demonstrates that the employment of energy gradient theory in studies of instability in natural convection is feasible. And the relationship between K and flow instability in a temperature field reveals the physical mechanism of flow instability of natural convection.

Furthermore, Saha et al. [32] validated the conclusions of scaling analysis by making some simulations of unsteady natural convection. They found that the boundary layer consists of three sub-layers. They are inner viscous layer, viscous layer and thermal layer respectively. It can be seen in Fig. 7(a) that the cold boundary layer above the bottom can be divided into three layers. This result is consistent with the conclusions of Saha et al [32].

**5.2 Effect of fin number on heat transfer**

Figures 8-9 show the contours of the temperature and the value of K with different fin number respectively, and Ra is $3.38 \times 10^6$ here. Figures 10-11 show the contours of the temperature and the value of K with different fin number respectively, but the Rayleigh number Ra is $3.38 \times 10^9$. And the Nusselt number Nu of all the numerical results are shown in Table 2.

From the second row of Table 2 where Ra is $3.38 \times 10^6$, it is observed that the most efficient heat transfer occurs when there is no fin in the cavity. However, when the fin number is greater than or equal to one, effect of the fin number on heat transfer is negligible, for the largest Nusselt number difference is 0.12, which is negligible comparing to the lowest Nusselt number. It is observed from the contours of the value of K in Fig. 9 that the largest intensity of flow instability occurs when there is no fin. And this observation accords well with the results in Table 2. Thus, summary can be made as follows from Table 2 and Figs. 8-9. Firstly, the fin(s) blocks heat transfer when Ra is $3.38 \times 10^6$. Secondly, the effect of fin numbers on heat transfer is negligible when the fin number is grater or equal to one with the same Ra.

From the third row of Table 2 where Ra is $3.38 \times 10^9$, it is observed that the most efficient heat transfer occurs when there is one fin in the cavity. And it can be found that the heat transfer rate in the cavity where the fin number is greater or equal to one is much larger than that there is no fin in the cavity. Furthermore, when the fin number exceeds one, it is observed that the effect of fin number on heat transfer is negligible and the heat transfer is depressed comparing to that when there is one fin. Both from Fig. 10 and Fig. 11, it is observed that the intensity of flow instability is very weak when there is no fin in the cavity. Absolutely, the fin triggers flow instability from Figs. 10-11 and heat transfer will be enhanced. And it is observed that the base flow tends to flow in a laminar manner in the center of the cavity when the fin number exceeds one. In other word, the intensity of flow instability in the cases where the fin number exceeds one is weaker than that there is only one fin in the cavity. Hence, it is easy to make some conclusions as follows. Firstly, the fin(s) enhances heat transfer when Ra is $3.38 \times 10^9$. Secondly, the most efficient heat transfer occurs when there is one fin. Thirdly, when the fin number exceeds one, the effect of fin number on heat transfer is negligible, while the heat transfer rate decreases comparing to that when there is one fin in the cavity.

Patters and Imberger [6] pointed out that there is a relationship between Ra, Pr and A. When Ra is relatively small, heat transfer is dominated by conduction. However, convection will play a leading role in comparison to conduction as Ra increases gradually. It is observed from Table 2 that whether the fin blocks or enhances heat transfer depends on Ra. The conclusions of Patterson and Imberger can reveal the reason why the effect of fin on heat transfer depends on Ra. Moreover, when the fin number exceeds one with a relatively large Ra, the heat transfer rate is depressed comparing to that when there is one fin in the cavity. We will explain the reason in Section 5.3. Therefore, some conclusions can be drawn as follows from the conclusions in Ref. [6] and the numerical results in Table 2. Firstly, when Ra is relatively small where the heat transfer is dominated by conduction, the fin(s) blocks heat transfer, and the effect of fin number on heat transfer is negligible when the fin number is greater or larger then one. Secondly, when Ra is relatively large where the heat transfer is dominated by convection, the fin(s) enhances heat transfer, and the effect of fin number on heat transfer is negligible as the fin number exceeds one. Furthermore, the heat transfer rate decreases when the fin number exceeds one comparing to that when there is only one fin in the cavity.

**5.3 Effect of the fin position on heat transfer**

Figures 12-13 show the temperature contours and the contours of the value of K, and Ra is $3.38 \times 10^9$. It can be seen in these two figures that the fin is fixed on different height of the hot wall. All the numerical results are achieved at 60s with the same governing parameters in order to research the effect of the fin position on heat transfer. Table 3 shows the numerical results achieved with the same Ra related in this section, with which the heat transfer is dominated by convection.

It can be seen from Table 3 that the most efficient heat transfer occurs when the fin is fixed at the middle height of the hot wall. Except for that case, the effect of fin position on heat transfer is negligible, for the Nusselt number varies in a very small range. It can also be observed from Figs. 12-13 that the base flow is stratified into so many layers when the fin is fixed at the middle height of the cavity, and the base flow is most likely to flow in a turbulent manner. This observation is in good accordance with the data in Table 3.

Saha et al. [32] pointed out that the boundary layer of the thermal flow can be divided into three sub-layers, and the three sub-layers develop separately. In the present study, the heat flux is accumulated along the hot wall with the development of the three sub-layers. And, the total heat flux reaches its maximum when the three sub-layers focus on the middle height of the hot side. Thus, when the fin is fixed at the middle height of the cavity with a relatively large Ra, the flow of the heat flux along the hot side is easier to lose its stability due to the fin. When the fin is not fixed at the middle height of the cavity, the fin still can trigger flow instability, but the intensity of the flow instability is much weaker than that when the fin is fixed at the middle height of the cavity. Hence, the most efficient heat transfer occurs when the fin is fixed at the middle height of the cavity with a relatively large Ra.

In section 5.2, it is found that the heat transfer rate with the fin number exceeding one is depressed comparing to that when there is only one fin. It can be explained as follow. When the fin number exceeds one, the development of three sub-layers is affected, which in turn hinders the accumulation of the heat flux. Thus, the intensity of flow instability is much weaker than that when the fin is fixed at the middle height of the hot side.

**5.4 Effect of fin length on heat transfer**

The effect of fin length on heat transfer is discussed also in this paper. The following numerical results in this section are still achieved at 60s with the same governing parameters. The fin length varies from 25mm to 75mm and the variation range of the fin length is small comparing to the length of the cavity L. Figures 14-15 show the temperature contours and the contours of the value of K. Table 4 shows the numerical results achieved with $Ra = 3.38 \times 10^9$, with which the heat transfer is dominated by convection.

It can be seen in Table 4 that the Nusselt number varies within a small range, which is negligible comparing to the average Nusselt number when the fin length varies from 25mm to 75mm. It is also can be examined from Figs. 14-15 that the intensity of flow instability is almost the same. And this observation in Figs. 14-15 accords well with the data in Table 4. It can be learned in Ref. [13] that the development of the boundary layer on the hot side in a differentially heated cavity depends on the fin length. And, during the transition of boundary layer on the hot side, the oscillatory will be produced which triggers heat transfer. When the fin is short, the horizontal intrusion under the fin will directly reattach to the hot side downstream of the fin after bypassing the fin. Thus, the oscillatory produced travels along the hot side downstream. However, if the fin length is long in comparison to a short one within a certain range, the horizontal intrusion under the fin will change into a starting plume and entrains into the boundary layer under the ceiling of the cavity. And the oscillatory will be produced under the ceiling of the cavity, which travels a shorter distance compared with the case with a short fin. However, as long as the flow time increases to a certain extent, the flow of the boundary layer on the hot side in all the above cases tends to be stable. The effect of migration distance of oscillatory tends to be slight. Thus, the effect of the fin length varying from 25mm to 75mm on heat transfer is not distinctive when heat transfer is dominated by convection.

**5.5 Correlation between Ra and $K_{max}$**

Figure 16 shows the correction between the Rayleigh number, Ra, and the maximum of the energy gradient function, $K_{max}$. It should be noted that the values on x-coordinate and y-coordinate are logarithm functions with the base of $\sqrt{10}$. Here, $K_{max}$ realizes the intensity of flow instability of a position in a flow field based on energy gradient theory, which means that the base flow is more unstable with a higher value of K. And, Ra represents the intensity of convection of natural convection based on the conclusions of Patterson and Imberger [6]. It is observed in Fig. 16 that the fit curve of correction between Ra and $K_{max}$ tends to be linear. This result is not accidental, and there must an inherent relationship between Ra and $K_{max}$.

It is known from the conclusions of Patterson and Imberger [6] that the convection will be enhanced when Ra increases gradually, which result in a more effective heat transfer. It is also well known that the heat transfer will be enhanced when the base flow of natural convection flows in turbulent manner. While $K_{max}$ represents the highest energy gradient at one position in a flow field, with which the base flow tends to lose its stability and flows in a turbulent manner. Thus, heat transfer will be enhanced with a higher $K$. Hence, the promotion of heat transfer is able to be characterized by Ra or $K_{max}$. The relationship between Ra and $K_{max}$ can be explained as follows. When Ra increases gradually, convection will play a leading role in comparison to conduction, which triggers flow instability of transient natural convection characterized by the increase of $K_{max}$ and heat transfer. The relationship between $K_{max}$ and Ra in Fig.16 is obtained for natural convection heat transfer problem in this study. For other flow and heat transfer problem, it needs to be further studied in future to clarify the universality.

### 5.6 Effect of Ra on heat transfer

Corcione [33] stated that the heat transfer rate from any heated or cooled boundary surface of the enclosure increases as the Rayleigh number Ra increases. Figure 17 shows the correction between Nu and Ra. And the numerical results are achieved at 60s with all the same governing parameters except Ra. It can be seen in Fig. 17 that the Nusselt number increases with the increase of Ra. And this result accords well with the conclusion of Corcione [33].

It is known from the conclusion of Patterson and Imberger [6] that whether the heat transfer is dominated by conduction or convection depends on Ra. As the former analysis between $K_{max}$ and Ra, the base flow tends to lose its stability and flow in a turbulent manner with the increase of Ra. Furthermore, when Ra is relatively small, the fin hinders heat transfer, however, the result is opposite with a relatively large Ra comparing to that when there is no fin in the cavity. Thus, $K_{max}$ and Ra reveal the same physical mechanism of natural convection from different perspective.

# 6 Conclusions

In this paper, natural convection flow in a differentially heated cavity is numerically investigated. All the numerical procedures are based on the unsteady Navier-Stokes equations and Boussinesq approximation. The energy gradient theory is employed to study the flow instability. The effects of fin number, fin position, and fin length on heat transfer in natural convection are analyzed. The correlation between $K_{max}$ and Ra and the effect of Ra on heat transfer are discussed. The conclusions are summarized as follows:

(1) The positions where flow instabilities occur accord well with the positions with the higher value of K, which in turn demonstrates that the application of energy gradient theory in unsteady natural convection is reliable.

(2) The criteria of flow instability based on energy gradient theory reveals the physical mechanism of flow instability in natural convection.

(3) When Ra is relatively small, the fin(s) blocks the heat transfer. The effect of the fin number on heat transfer is negligible when the fin number is larger than one.

(4) When Ra is relatively large, the fin(s) enhances heat transfer. The effect of fin number on heat transfer is also negligible when the fin number exceeds one. However, the heat transfer rate decreases when the fin number exceeds one comparing to that when there is only one fin in the cavity.

(5) Most efficient heat transfer occurs when the fin is fixed at the middle height of the cavity with a relatively large Ra, which is due to the development of the three sub-layers.

(6) The effect of the fin length varying from 25mm to 75mm on heat transfer is not distinctive when heat transfer is dominated by convection.

(7) Convection will play a leading role gradually with the increase of Ra, which triggers flow instability of natural convection characterized by the increase of $K_{max}$. Thus, the correlation between $K_{max}$ and Ra tends to be linear.

(8) As the Nusselt number Nu increases with the increase of Rayleigh number Ra, heat transfer is enhanced. The maximum of the energy gradient function, $K_{max}$, and the Rayleigh number Ra reveals the same physical mechanism of natural convection from a different perspective.

## Acknowledgment

This work is supported by the National Natural Science Foundation of China (51579224, 51536008) and the Zhejiang Province Key Science and Technology Innovation Team Project (2013TD18). The authors thank Dr. Chengwei Lei in The University of Sydney for the helpful discussions.

References
[1] A. E. Gill, The boundary-layer regime for convection in a rectangular cavity, J. Fluid Mech., 26, 515-536, 1966.
[2] G. De Vahl Davis, Natural convection of air in a square cavity: a bench mark numerical solution, Int. J. Numer. Meth. Fluids, 3, 249-264, 1983.
[3] V. A. Akinsete, T. A. Coleman, Heat transfer by steady laminar free convection in triangular enclosures, Int. J. Heat Mass Transfer, 25, 991-998, 1982.

[4] I. Dagtekin, H. F. Oztop, J. P. Hartnett, and W. J. Minkowycz, Natural convection heat transfer by heated partitions within enclosure, Int. Comm. Heat Mass Transfer, 28(6), 823-834, 2001.

[5] R. B. Yedder, E. Bilgen, Laminar natural convection in inclined enclosures bounded by a solid wall, Heat and Mass Transfer, 32, 455-462, 1997.

[6] J. C. Patterson, J. Imberger, Unsteady natural convection in a rectangular cavity, J. Fluid Mech., 100, 65-86, 1980.

[7] C. Y. Soong, P. Y. Tzeng, D. C. Chiang, T. S. Sheu, Numerical study on mode-transition of natural convection in differentially heated inclined enclosures, Int. J. Heat Mass Transfer, 39(14), 2869-2882, 1996.

[8] H. Asan, L. Namli, Laminar natural convection in a pitched roof of triangular cross-section: summer day boundary conditions, Energy and Buildings, 33, 69-73, 2009.

[9] F. Corvaro, M. Paroncini, An experimental study of natural convection in a differentially heated cavity through a 2D-PIV system, International Journal of Heat and Mass Transfer, 52, 355-365, 2009.

[10] M. Rokni, B. Sunden, Investigation of a two-equation turbulent heat transfer model applied to ducts, J. Heat Transfer-Trans. ASME, 125 (1), 194-200, 2003.

[11] F. Xu, J. C. Patterson, C. Lei, An experimental study of the unsteady thermal flow around a thin fin on a differentially heated cavity, International Journal of Heat and Fluid Flow, 29, 1139-1153, 2008.

[12] F. Xu, J. C. Patterson, C. Lei, Transition to a periodic flow induced by a thin fin on the sidewall of a differentially cavity, International Journal of Heat and Mass Transfer, 52, 620-628, 2009.

[13] F. Xu, J. C. Patterson, C. Lei, Transient natural convection flows around a thin fin on the sidewall of a differentially heated cavity, J. Fluid Mech. 639, 261-290, 2009.

[14] F. Xu, J. C. Patterson, C. Lei, Temperature oscillations in a differentially heated cavity with and without a fin on the sidewall, International Communications in Heat and Mass Transfer, 37, 350-359, 2010.

[15] F. Xu, J. C. Patterson, C. Lei, Unsteady flow and heat transfer adjacent to the sidewall of a differentially heated cavity with a conducting and an adiabatic fin, International Journal of Heat and Fluid Flow, 32, 680-687, 2011.

[16] E. Bilgen, Natural convection in cavities with a thin fin on the hot wall, International Journal of Heat and Mass Transfer, 48, 3493-3505, 2005.

[17] H. Oztop, E. Bilgen, Natural convection in differentially heated and partially divided square cavities with internal heat generation. International Journal of Heat and Fluid Flow, 27, 466-475, 2006.

[18] A. Mezrhab, M. Jami, C. Abid, M. Bouzidi, P. Lallemand, Lattice-Boltzmann modeling of natural convection in an inclined square enclosure with partitions attached to its cold wall, International Journal of Heat and Fluid Flow, 27, 456-465, 2006.

[19] Y. Varol, H. F. Oztop, A. Varol, Effects of thin fin on natural convection in porous triangular enclosures, International Journal of Thermal Sciences, 46, 1033-1045, 2007.

[20] A. H. Mahmoudi, M. Shahi, A. H. Raouf, A. Ghasemian, Numerical study of natural convection cooling of horizontal heat source mounted in a square cavity filled with nanofluid, International Communications in Heat and Mass Transfer, 37, 1135-1141, 2010.

[21] H.-S. Dou and N. Phan-Thien, Instability of Fluid Material Systems, 15th Australasian Fluid Mechanics Conference, The University of Sydney, Sydney, Australia 13-17 December, 2004.

[22] H.-S. Dou, Viscous instability of inflectional velocity profile, Proceedings of the Forth International Conference on Fluid Mechanics, Tsinghua University Press &Springer-Verlag, 2004.

[23] H.-S. Dou, Mechanism of flow instability and transition to turbulence, International Journal of Non-Linear Mechanics, 41, 512-517, 2006.

[24] H.-S. Dou, Three important theorems for flow stability, Proceedings of the Fifth International Conference on Fluid Mechanics, Tsinghua University Press & Springer,

2007.
[25] H.-S. Dou and B. C. Khoo, Mechanism of wall turbulence in boundary layer low, Modern Physics Letters B., 23(3), 457-460, 2009.
[26] H.-S. Dou and B. C. Khoo, Criteria of turbulent transition in parallel flows, Model Physics Letters B., 24(13), 1437-1440, 2010.
[27] H.-S. Dou and B. C. Khoo, Investigation of Turbulent transition in plane Couette Flows Using Energy Gradient Method, Advances in Appl. Math, Mech., 3(2), 165-180, 2011.
[28] H.-S. Dou and B. C. Khoo, Energy Loss Distribution in the Plane Couette Flow and the Taylor-Couette Flow between Concentric Rotating Cylinders, Int. J. Thermal Sci., 46, 262-275, 2007.
[29] H.-S. Dou and B. C. Khoo, Instability of Taylor-Couette Flow between Concentric Rotating Cylinders, Int. J. Thermal Sci., 47, 1422-1435, 2008.
[30] H.-S. Dou, G. Jiang and C. Lei, Numerical Simulation and Stability Study of Natural Convection in an Inclined Rectangular Cavity, Mathematical Problems in Engineering, 10.1155/2013/198695
[31] G. K. Bachelor, Heat transfer by free convection across a closed cavity between vertical boundaries at different temperature, Q. Appl. Math. 12, 209-233, 1954.
[32] S. C. Saha, J. C. Patterson, C. Lei, Scaling of Natural Convection of an Inclined Flat Plate: Sudden Cooling Condition, Journal of Heat Transfer, 133 / 041503-1, 2011.
[33] M. Corcione, Effects of the thermal boundary conditions at the sidewalls upon natural convection in rectangular enclosures heated from below and cooled from above, International Journal of Thermal Sciences, 42, 199-208, 2003.

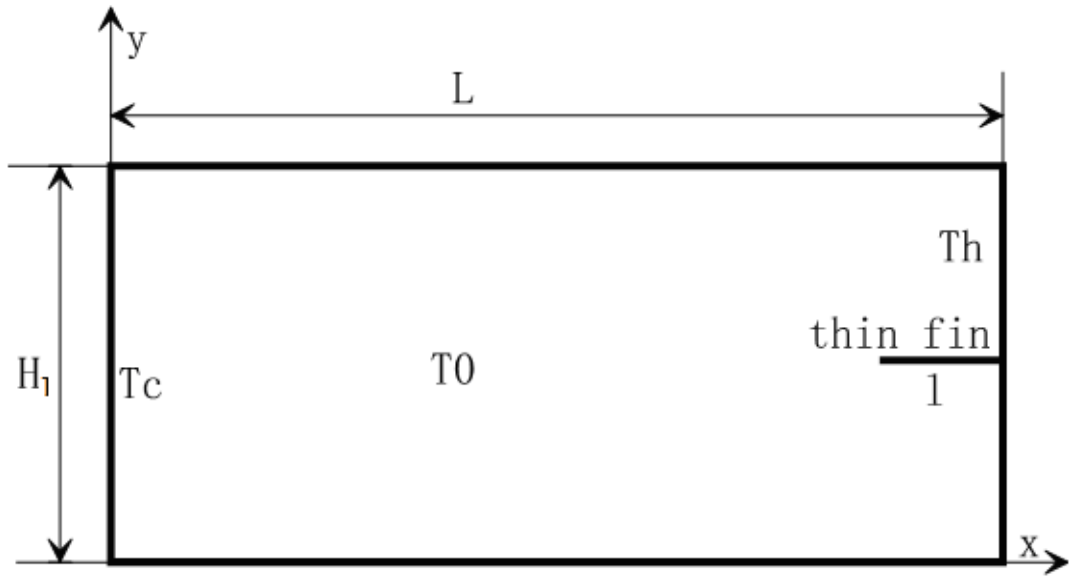

Figure 1:   Numerical geometry

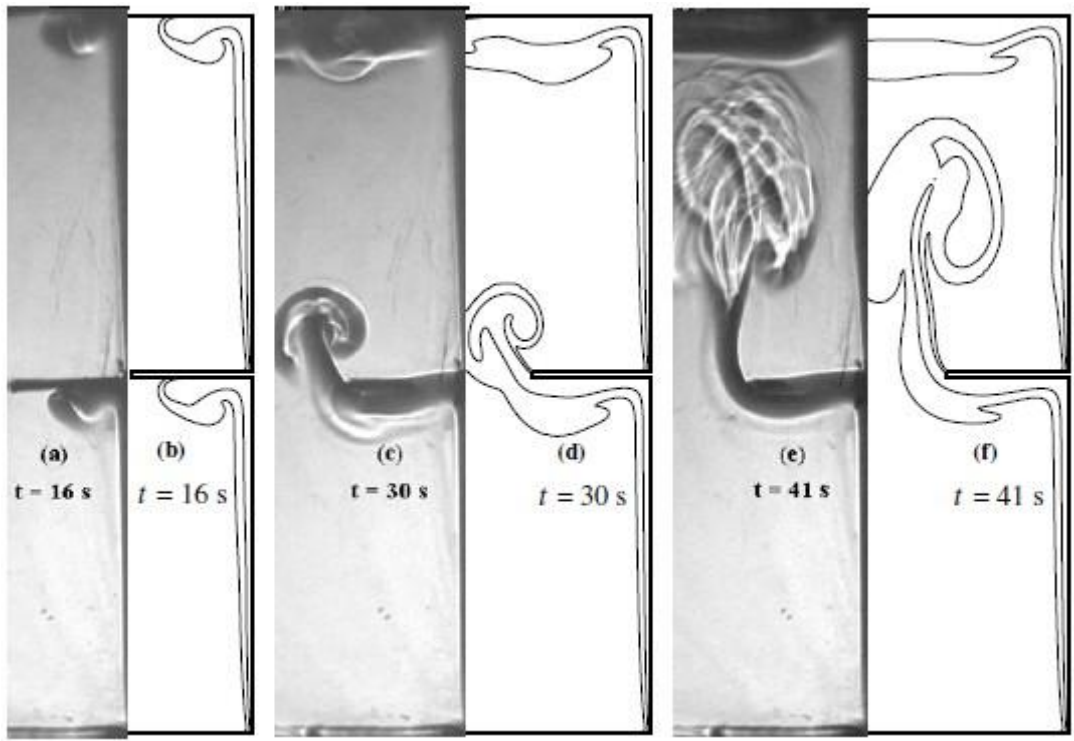

Figure 2: Temperature distribution at different time (Ra=$1.84 \times 10^9$) in [12].

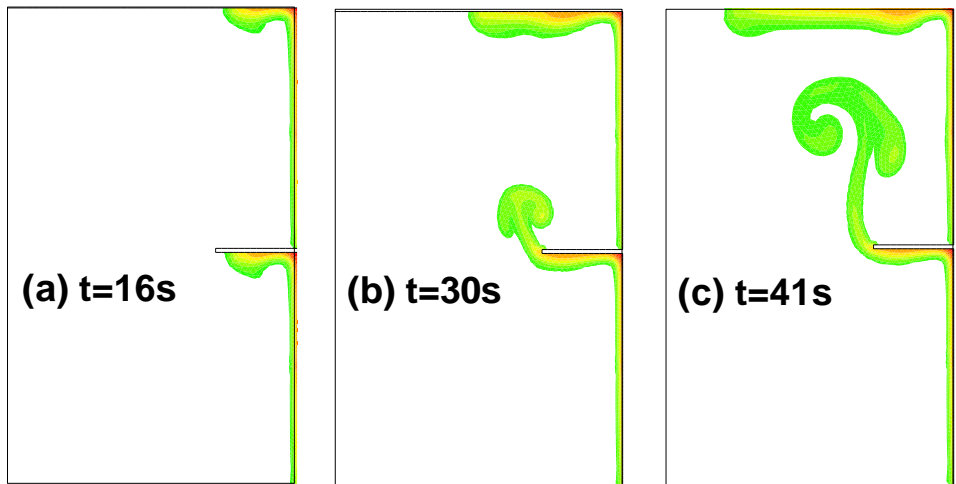

Figure 3: Present numerical results (Ra=$1.84 \times 10^9$).

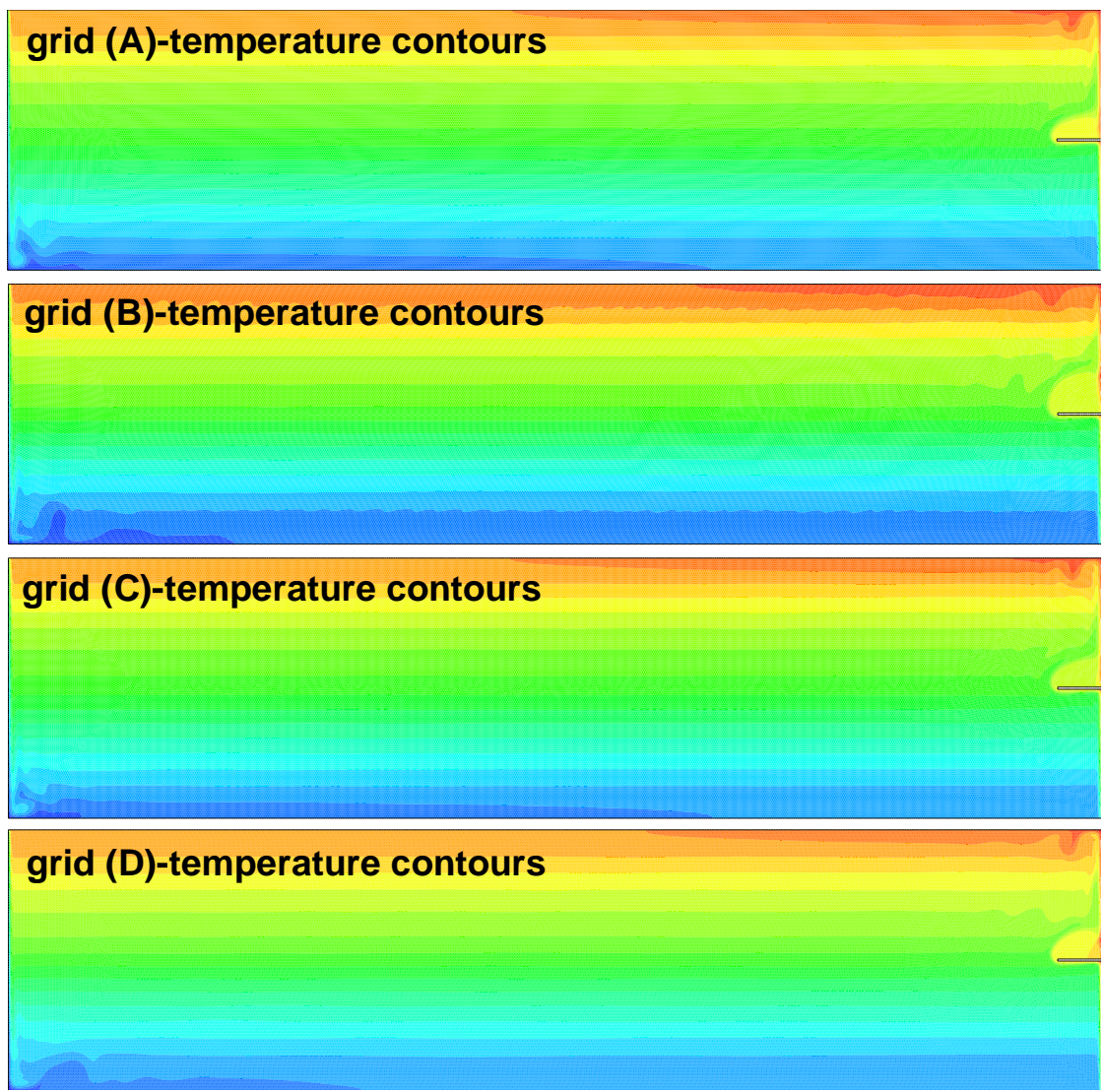

Figure 4: Temperature contours with four different grid meshes: gird (A) $\Delta x = 0.004$, grid (B) $\Delta x = 0.003$, grid (C) $\Delta x = 0.002$ and grid (D) $\Delta x = 0.0015$.

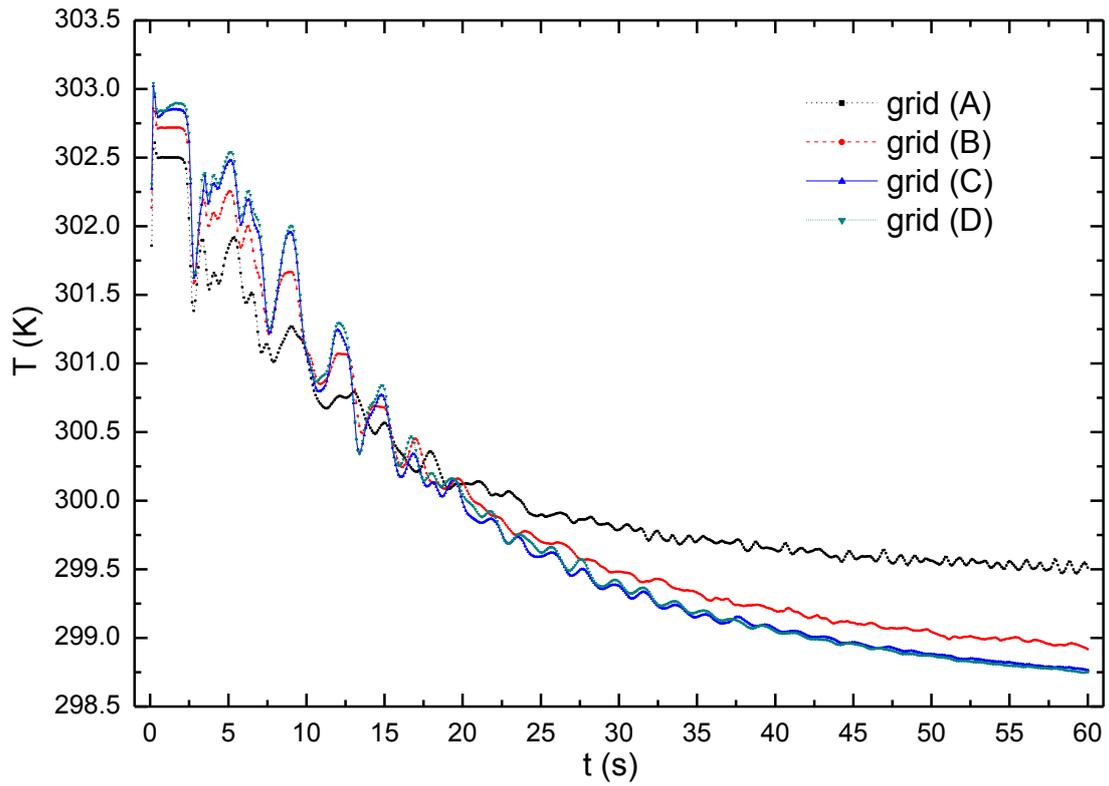

Figure 5: Temperature profile calculated with four different grid meshes: gird (A) $\Delta x = 0.004$, grid (B) $\Delta x = 0.003$, grid (C) $\Delta x = 0.002$ and grid (D) $\Delta x = 0.0015$.

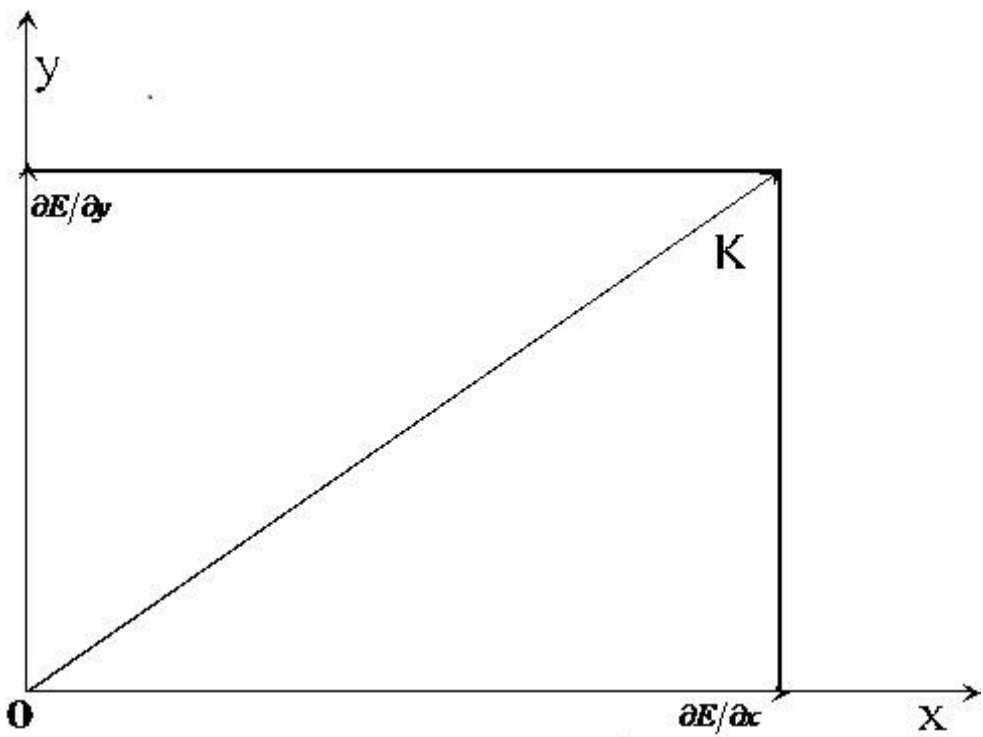

Figure 6: Calculation of the value of K

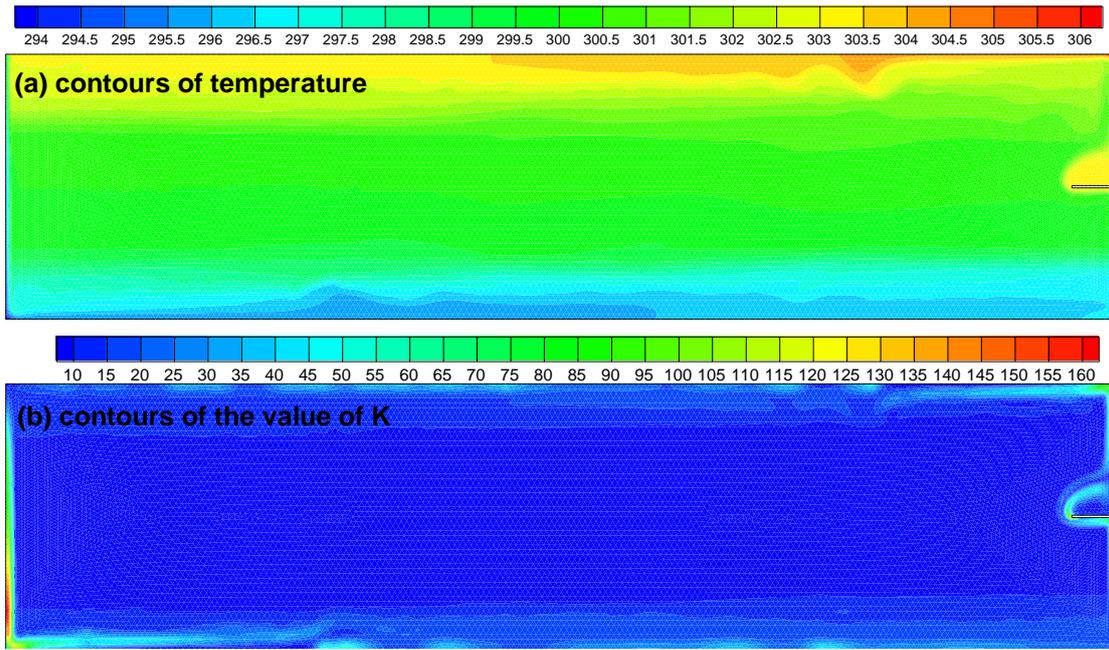

Figure 7: Validation of the criteria of flow instability based on energy gradient method: (a) contours of temperature, and (b) contours of the value of K

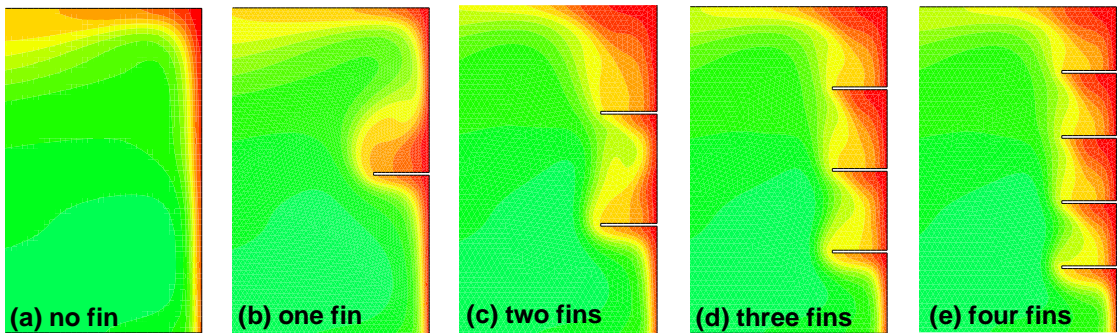

Figure 8: Temperature contours with different fin number fixed in the cavity, and Ra is $3.38 \times 10^6$: (a) no fin, (b) one fin, (c) two fins, (d) three fins, and (e) four fins.

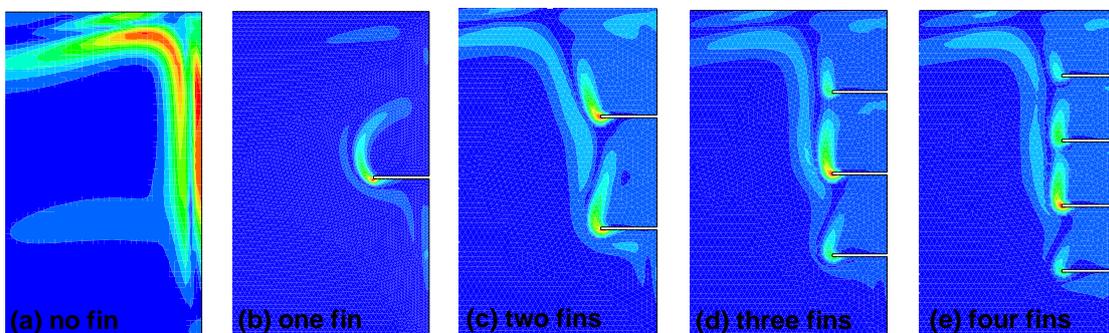

Figure 9: Contours of the value of K with different fin number fixed in the cavity, and Ra is $3.38 \times 10^6$: (a) no fin, (b) one fin, (c) two fins, (d) three fins, and (e) four fins.

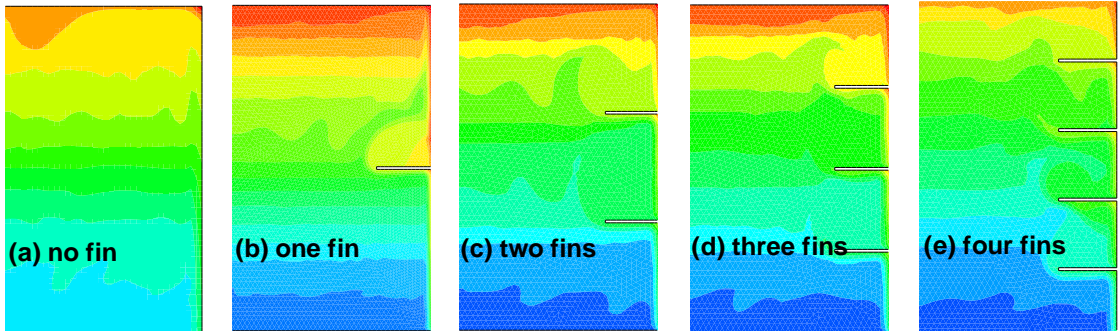

Figure 10: Temperature contours with different fin number fixed in the cavity, and Ra is $3.38 \times 10^9$: (a) no fin, (b) one fin, (c) two fins, (d) three fins, and (e) four fins.

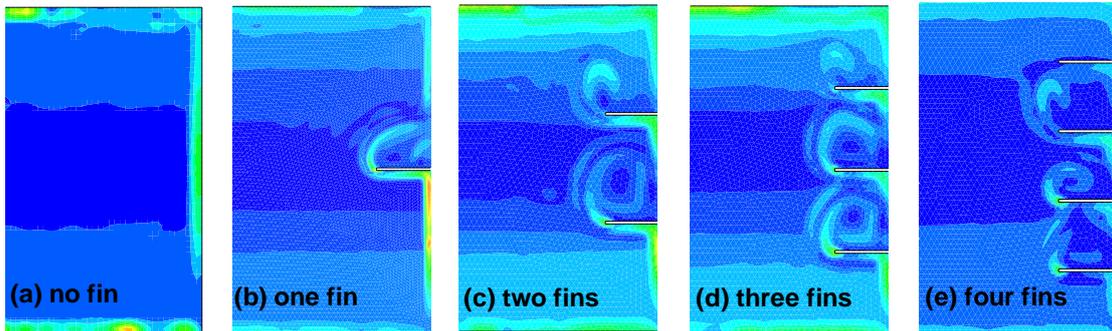

Figure 11: Contours of the value of K with different fin number fixed in the cavity, and Ra is $3.38 \times 10^9$: (a) no fin, (b) one fin, (c) two fins, (d) three fins, and (e) four fins.

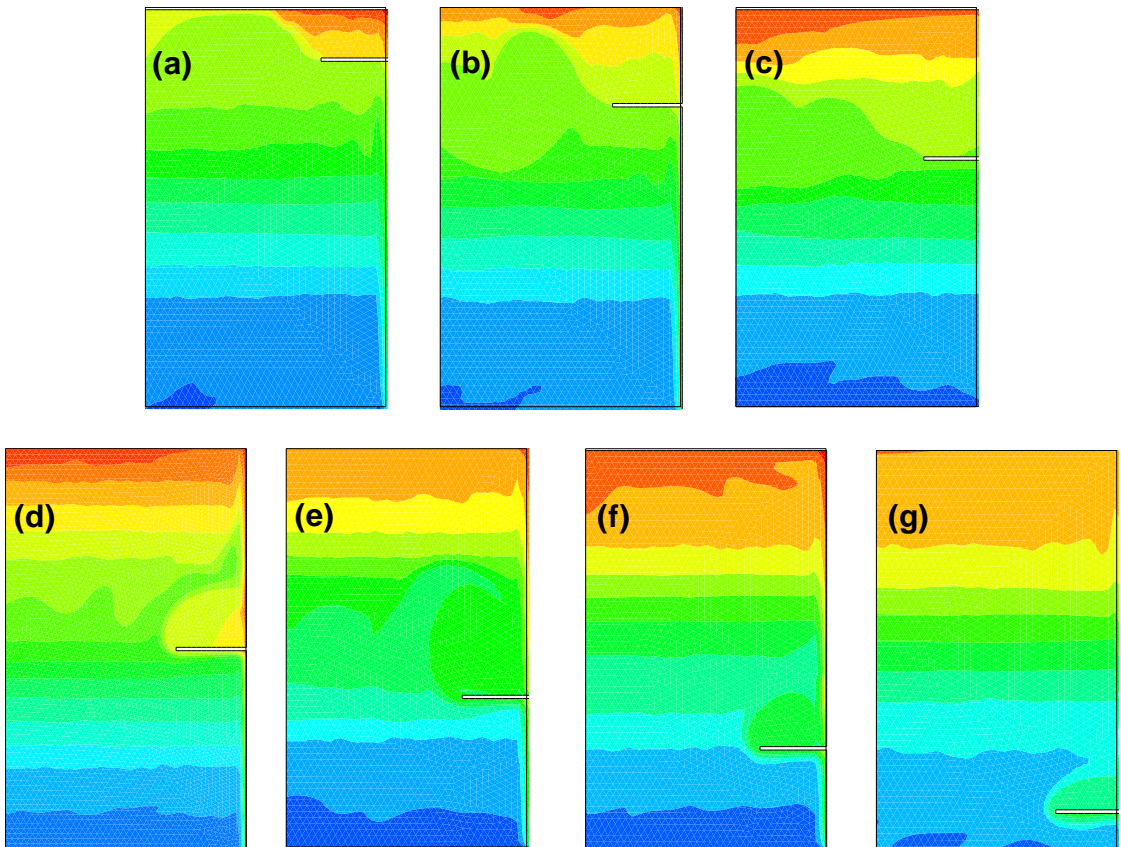

Figure 12: Temperature contours with different fin position, and Ra is $3.38 \times 10^9$: height of the fin, (a) 0.21m, (b) 0.18m, (c) 0.15m, (d) 0.12m, (e) 0.09m, (f) 0.06m, (g) 0.03m.

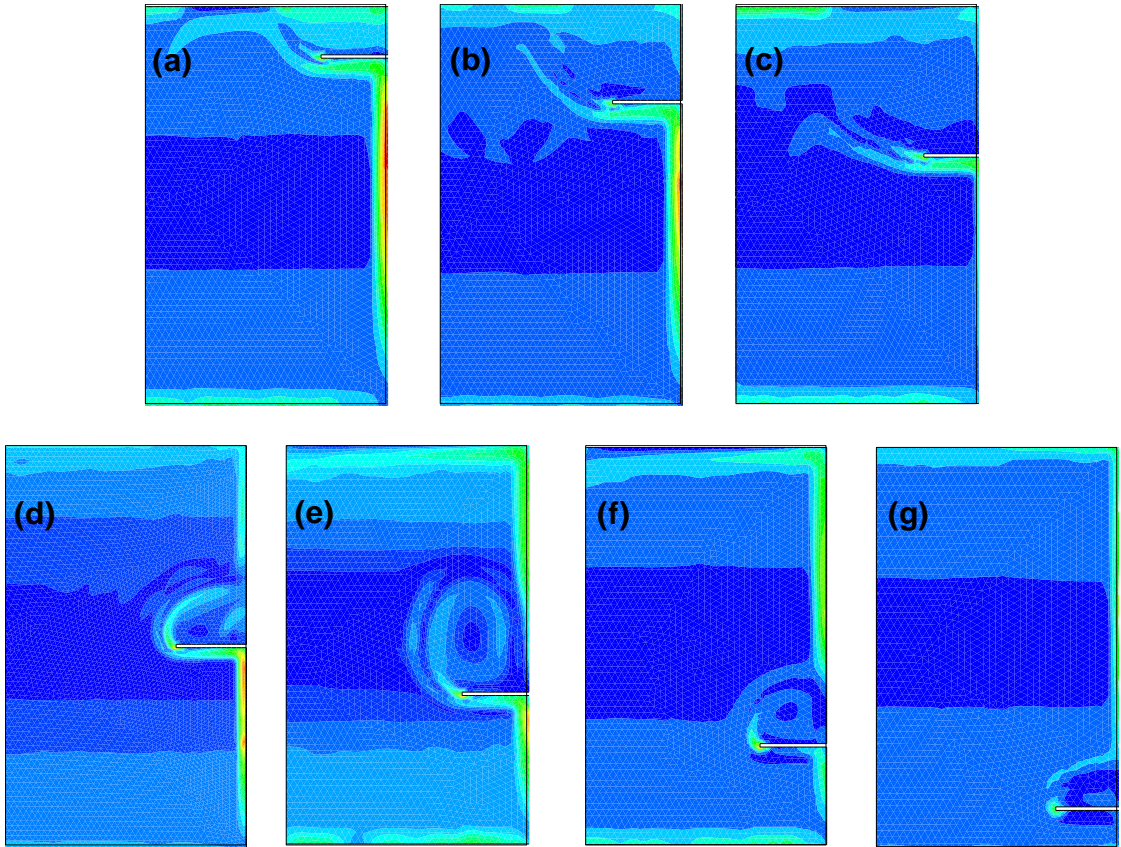

Figure 13: Contours of the value of K with different fin position, and Ra is $3.38 \times 10^9$: height of the fin, (a) 0.21m, (b) 0.18m, (c) 0.15m, (d) 0.12m, (e) 0.09m, (f) 0.06m, (g) 0.03m.

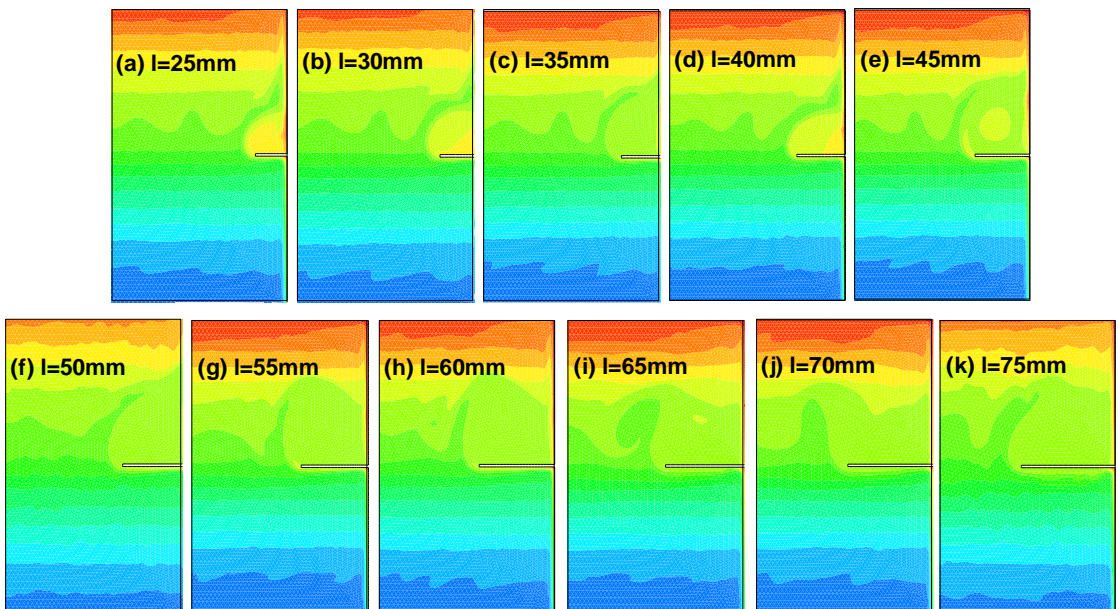

Figure 14: Temperature contours with different fin length, and Ra is $3.38 \times 10^9$.

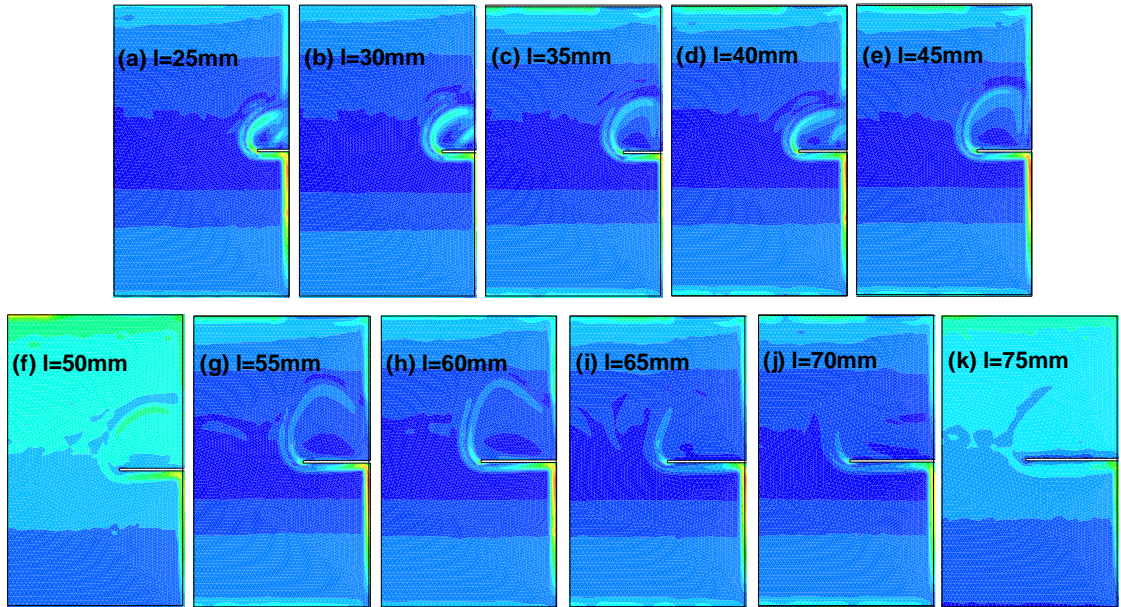

Figure 15: Contours of the value of K with different fin length, and Ra is $3.38 \times 10^9$.

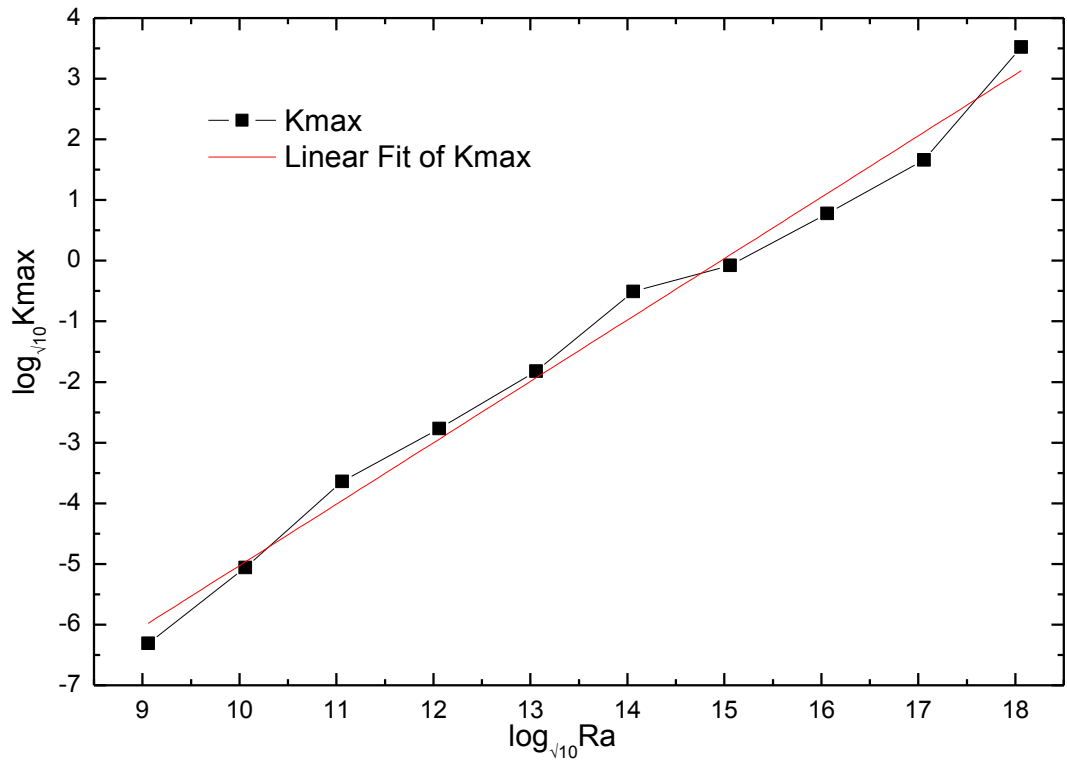

Figure 16: Fit curve between Ra and $K_{max}$

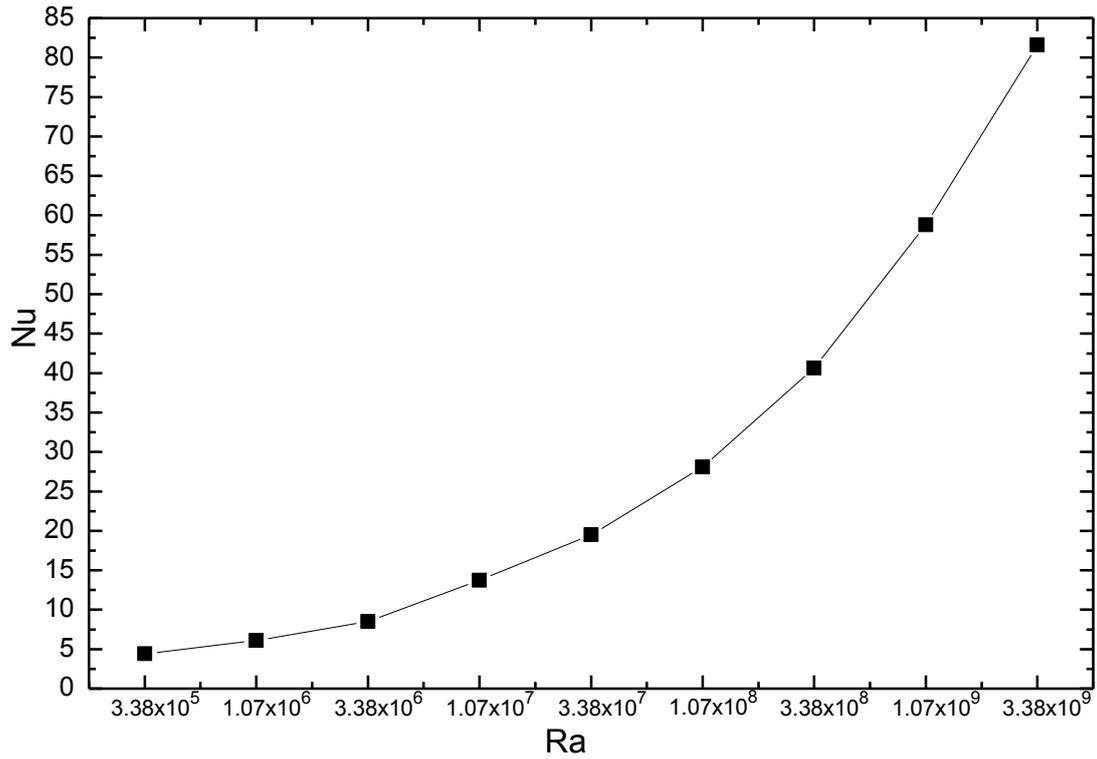

Figure 17: Numerical results of Nu versus Ra

Table 1: Comparison of positions of the thermal flow plume fronts on time at Ra=$1.84\times10^9$ (height of plume/height of cavity)

| Methods | t=30 s | t=41 s |
|---|---|---|
| Experiment in [12] | 0.1250 | 0.4166 |
| Simulations in [12] | 0.1250 | 0.3542 |
| Simulations in present work | 0.1250 | 0.3549 |

Table 2: Effect of fin number on heat transfer

| Ra | No fin (Nu) | One fin (Nu) | Two fins (Nu) | Three fins (Nu) | Four fins (Nu) |
|---|---|---|---|---|---|
| $3.38\times10^6$ | 9.24 | 8.48 | 8.60 | 8.56 | 8.54 |
| $3.38\times10^9$ | 52.48 | 81.60 | 66.18 | 65.32 | 65.58 |

Table 3: Effect of the fin position on heat transfer

| | (a) | (b) | (c) | (d) | (e) | (f) | (g) |
|---|---|---|---|---|---|---|---|
| Nu | 66.68 | 66.90 | 67.02 | 81.6 | 67.0 | 67.3 | 68.4 |

Table 4: Effect of fin length on heat transfer

| Fin length (mm) | 25 | 30 | 35 | 40 | 45 | 50 | 55 | 60 | 65 | 70 | 75 |
|---|---|---|---|---|---|---|---|---|---|---|---|
| Nu | 81.8 | 82.9 | 82.2 | 81.6 | 82.4 | 78.9 | 82.3 | 82.2 | 82.1 | 82.5 | 78.9 |